\documentclass[11pt,amssymb]{article}
\usepackage{url}
\usepackage{hyperref}
\usepackage{breakurl}
\newcommand{\lbrk}{{\linebreak[0]}}
\newcommand{\less}{{\;\scriptstyle\stackrel{<}{=}\;}}
\newcommand{\more}{{\;\scriptstyle\stackrel{>}{=}\;}}

\newtheorem{theorem}{Theorem}[section]
\newtheorem{definition}{Definition}[section]
\newtheorem{lemma}{Lemma}[section]
\newtheorem{corollary}{Corollary}[section]
\newtheorem{conjecture}{Conjecture}[section]

\newcommand{\arsin}{\rm arsin}
\newcommand{\arcos}{\rm arcos}
\newcommand{\artan}{\rm artan}
\newcommand{\arsinh}{\rm arsinh}
\newcommand{\arcosh}{\rm arcosh}
\newcommand{\artanh}{\rm artanh}

\newcommand{\W}{\hphantom{0}}

\begin{document}

\vspace*{12mm}
\begin{center}
{\bf THE COMPLEXITY OF MULTIPLE-PRECISION}\\[2ex]
{\bf ARITHMETIC}
\footnote{First appeared in 
{\em The Complexity of Computational Problem Solving}
(edited by R S Anderssen and R P Brent), Univ.~of Queensland Press,
1976, 126--165.
Retyped with minor corrections by
Frances Page at Oxford University Computing
Laboratory, 1999.\\
Copyright \copyright\ 1976, 1999, R.~P.~Brent and
University of Queensland Press.
\hspace*{\fill} rpb032 typeset using \LaTeX.}\\[6ex]

{\bf Richard P Brent}\\[6ex]

{\bf Computer Centre, Australian National University}
\end{center}
\vspace*{8mm}

\begin{quote}
In studying the complexity of iterative processes it is usually assumed
that the arithmetic operations of addition, multiplication, and division
can be performed in certain constant times.  This assumption is invalid
if the precision required increases as the computation proceeds.  We give
upper and lower bounds on the number of single-precision operations
required to perform various multiple-precision operations, and deduce
some interesting consequences concerning the relative efficiencies of
methods for solving nonlinear equations using variable-length
multiple-precision arithmetic.
\end{quote}

\vspace*{4mm}

\section{Introduction}
\thispagestyle{empty}

Traub $[28]$ defines analytic computational complexity to be the optimality
theory of analytic or continuous processes.  Apart from some work by
Schultz $[24]$ on differential equations, most recent results have
concerned iterative methods for the solution of nonlinear equations or
systems of equations.  See, for example, Brent $[$1,3,6$]$, Brent,
Winograd and Wolfe $[7]$, Kung $[14,15]$, Kung and Traub $[$16--17$]$,
Paterson $[21]$, Rissanen $[22]$, Traub $[$28--32$]$ and Wozniakowski
$[35, 36]$.\\

The authors just cited make the (usually implicit) assumption that
arithmetic is performed with a fixed precision throughout a given 
computation.  This is probably true for most computations programmed
in Fortran or Algol 60.  Suppose, though, that we are concerned with
an iterative process for approximating an irrational number $\zeta$
(for example, $\sqrt{2}$, $\pi$ or $e$) to arbitrary accuracy.
The iterative process should (theoretically) generate a sequence
($x_i$) of real numbers, such that $\displaystyle\zeta = \lim_{i \rightarrow \infty}
x_i$, provided no rounding errors occur.  On a computing machine
each $x_i$ has to be approximated by a finite-precision 
machine-representable number $\widetilde{x}_i$, and
$\displaystyle\zeta = \lim_{i \rightarrow \infty} \widetilde{x}_i$  
can only hold if the precision
increases indefinitely as $i \rightarrow \infty$.  In practice, only
a finite number of members of the sequence $(\widetilde{x}_i)$ will ever
be generated, but if an accurate approximation to $\zeta$ is required
it may be possible to save a large amount of computational work by
using variable precision throughout the computation.  This is likely
to become easier to program as new languages (and possibly hardware), 
which allow the precision of floating-point numbers to be varied
dynamically, are developed.\\

In Section 7 we discuss the effect of using variable precision when 
solving nonlinear equations.  Before doing so, we consider the complexity
of the basic multiple-precision arithmetic operations.  We assume
that a standard floating-point number representation is used, with
a binary fraction of $n$ bits.  (Similar results apply for any fixed
base, for example, 10.)  We are interested in the case where $n$ is
much greater than the wordlength of the machine, so the fraction 
occupies several words.  For simplicity, we assume that the exponent 
field has a fixed length and that numbers remain in the allowable 
range, so problems of exponent overflow and underflow may be neglected.
Note that our assumptions rule out exotic number representations (for
example, logarithmic $[4]$ or modular $[33, 34]$ representations) in
which it is possible to perform some (but probably not all) of the
basic operations faster than with the standard representation.  To
rule out ``table-lookup'' methods, we assume that a random-access
memory of bounded size and a bounded number of sequential tape units
are available.  (Formally, our results apply to multitape Turing 
machines.)\\

In Sections 2 to 6 we ignore ``constant'' factors, that is factors
which are bounded as $n \rightarrow \infty$.  Although the constant
factors are of practical importance, they depend on the computer and
implementation as well as on details of the analysis.  Certain 
machine-independent constants are studied in Sections 7 and 8.\\

If $B$ is a multiple-precision operation, with operands and result
represented as above (that is, ``precision $n$'' numbers), then
$t_n(B)$ denotes the worst-case time required to perform $B$, obtaining
the result with a relative error at most $2^{-n}c$, where $c$ is
independent of $n$.  We assume that the computation is performed on a 
serial machine whose single-precision instructions have certain constant 
execution times.  The following definition follows that in Hopcroft $[11]$.\\

\begin{definition}\rm
$B$ is {\it linearly reducible} to $C$ (written $B \less C$), if there
is a positive constant $K$ such that

$$t_n(B) \leq Kt_n(C) \eqno{(1.1)}$$\\[-4ex]

for all sufficiently large $n$.  $B$ is {\it linearly equivalent} to
$C$ (written $B \equiv C$) if $B \less C$ and $C \less B$.
\end{definition}

In Section 2 we consider the complexity of multiple-precision addition
and some linearly equivalent operations.  Then, in Section 3, we show that
multiple-precision division, computation of squares or square roots,
and a few other operations are linearly equivalent to multiplication.
Most of these results are well known $[8, 9]$.\\

Sections 4 and 5 are concerned with the ``operations'' of evaluating
exponentials, logarithms, and the standard trigonometric and hyperbolic
functions (sin, artan, cosh, and so on).  It turns out that most of 
(and probably all) these operations are linearly equivalent so long
as certain restrictions are imposed.\\

Section 6 deals with the relationship between the four equivalence
classes established in Sections 2 to 5, and several upper bounds on
the complexity of operations in these classes are given.  The best
known constants relating operations which are linearly equivalent to 
multiplication are given in Section 7.\\

Finally, in Section 8, we compare the efficiencies of various methods 
for solving nonlinear equations using variable-length multiple-precision
arithmetic.  The relative efficiencies are different from those for the 
corresponding fixed-precision methods, and some of the conclusions
may be rather surprising.  The results of Sections 4 to 8 are mainly
new.\\

In the analysis below, $c_1, c_2, \ldots\;$ denote certain positive
constants which do not need to be specified further.  The notation 
$f \sim g$ means that 
$\displaystyle\lim_{n \rightarrow \infty} f(n)/g(n) = 1$,
and $f = O(g)$ means that $|f(n)| \leq Kg(n)$ for some constant $K$
and all sufficiently large $n$.  Finally, the abbreviation ``{\it mp}''
stands for ``variable-length multiple-precision''.
  

\section{Addition and linearly equivalent operations}

Let $A$ denote the operation of multiple-precision addition.  Any
reasonable implementation of floating-point addition, using at least
one guard digit to avoid the possible occurrence of large relative
errors, gives

$$ t_n(A) \leq c_1n \;. \eqno{(2.1)}$$\\[-4ex]

Conversely, from the assumptions stated in Section 1, it is clear that

$$ t_n(A) \geq c_2n \;. \eqno{(2.2)} $$\\[-4ex]

Hence, the complexity of multiple-precision addition is easily
established.  (For the operations discussed in Sections 3 and 5 the
results are less trivial, in fact the conjectured lower bounds
corresponding to (2.2) have not been proved rigorously.)\\

It is easy to see that bounds like (2.1) and (2.2) hold for
multiple-precision subtraction, and multiplication or division of
a multiple-precision number by a single-precision number (or even by
any rational number with bounded numerator and denominator).  Hence,
all these operations are linearly equivalent  to addition.


\section{Multiplication and linearly equivalent operations}

Let $D, I, M, R$ and $S$ denote the multiple-precision operations 
of division, taking reciprocals, multiplication, extraction of square
roots and forming squares, respectively.  In this section, we show that
all these operations are linearly equivalent.  The proofs are
straightforward, but the result is surprising, as it seems intuitively
obvious that taking a square root is inherently ``more difficult''
than forming a square, and similarly for division versus multiplication.
(Some bounds on the relative difficulty of these operations are given
in Section 7.)\\

\begin{lemma}

$$ M \more S \more A \;. \eqno{(3.1)}$$
\end{lemma}

{\bf Proof.} Clearly
$$ t_n(M) \geq t_n(S) \geq c_3n \;, \eqno{(3.2)}$$\\[-4ex]

so the result follows from (2.1).\\

Sharp upper bounds on $t_n(M)$ are not needed in this section, so
we defer them until Section 6.  Lemmas 3.2 and 3.3, although weak,
are sufficient for our present purposes. \\

\begin{lemma}
For all positive n, 

$$ t_{2n}(M) \leq c_4t_n(M) \;. \eqno{(3.3)}$$
\end{lemma}

{\bf Proof.} First assume that $n$ is divisible by 3, and consider
operations on the $n$-bit fractions only.  If we can multiply $n$-bit
numbers with relative error $2^{-n}c_0$ then we can multiply
$n/3$-bit numbers exactly (assuming $2^{n/3} > 2c_0$).  Thus, a
$2n$-bit fraction $x$ may be split up into

$$ x = \lambda a + \lambda^2b + \ldots + \lambda^6f \;, \eqno{(3.4)}$$\\[-4ex]

where $\lambda = 2^{-n/3}$ and $a, b, \ldots, f$ are integers in
$\left[0,2^{n/3}\right)$, and the product of two such $2n$-bit fractions may
be formed exactly with 36 exact multiplications of $n/3$-bit
numbers and some additions.  Thus

$$ t_{2n}(M) \leq 36t_n(M) + c_5t_{2n}(A) \;, \eqno{(3.5)} $$\\[-4ex]

and the result follows from Lemma 3.1.  Trivial modifications to the
above proof suffice, if $n$ is not divisible by 3.\\

\begin{lemma}
For some constant $c_6<1$,

$$ t_n(M) \leq c_6t_{8n}(M) \eqno{(3.6)}$$

for all sufficiently large n.\\
\end{lemma}

{\bf Proof.}  If $a, b, c$ and $d$ are integers in $\left[0, 2^n\right)$, the
identity

$$ (a+\lambda b)(c+\lambda d)= ac + \lambda(bc+ad) + \lambda^2 bd\;,
\eqno{(3.7)}$$\\[-4ex]

with $\lambda = 2^{3n}$, may be used to obtain the products $ac$
and $bd$ from one $8n$-bit product.  Thus

$$ 2t_n(M) \leq t_{8n}(M) + c_7\;. \eqno{(3.8)}$$\\[-4ex]

The result (with $c_6 = 3/4$) follows if $n$ is sufficiently large
that $t_{8n}(M) \geq 2c_7$.  (We have assumed that the time required 
for one $n$-bit multiplication is half the time required for two
independent $n$-bit multiplications, but much weaker assumptions
would be sufficient.) \\

The following lemma will be used to estimate the work required
for multiple-precision divisions and square roots.\\

\begin{lemma}
Given $\alpha \in (0,1)$, there is a constant $c_8$ such that,
for any integers $n_0, \ldots, n_p$ satisfying

$$ 1 \leq n_j \leq \alpha^jn \eqno{(3.9)}$$\\[-4ex]

for $j=0, 1, \ldots, p$, we have

$$ \sum^{p}_{j=0} t_{n_j}(M) \leq c_8 t_n(M)\;. \eqno{(3.10)}$$\\[-4ex]
\end{lemma}

{\bf Proof.}  Let $k$ be large enough that

$$\alpha^k \leq 1/8\;. \eqno{(3.11)}$$\\[-4ex]

{From} (3.9) and (3.11),

$$ t_{n_{jk}}(M) \leq \alpha^{j}_{6} t_n(M) \eqno{(3.12)}$$\\[-4ex]

for $j=0, 1, \ldots, \lfloor p/k\rfloor$, provided $n_{jk}$ is
sufficiently large for Lemma 3.3 to be applicable.  Thus,

$$ \sum^{p}_{j=0} t_{n_j}(M) \leq kt_n(M)
               \left(1+ c_6+ c^2_6+\ldots\right)+c_7\;, \eqno{(3.13)}$$\\[-4ex]

where the term $c_7$ allows for those $t_{n_j}(M)$ for which 
Lemma 3.3 is not applicable.  If

$$ c_8 = k/\left(1-c_6\right) + c_7 \;,\eqno{(3.14)}$$\\[-4ex]

the result follows from (3.13).\\

The following lemma shows that multiple-precision multiplication is 
linearly equivalent to squaring.  This result is essentially due to 
Floyd  $[9]$.\\

\begin{lemma}

$$ M\equiv S\;. \eqno{(3.15)}$$\\[-4ex]
\end{lemma}

{\bf Proof.}  Since squaring is a special case of multiplication,

$$ M \more S\;. \eqno{(3.16)}$$\\[-4ex]

Conversely, we may use the identity

$$ 4\lambda ab = (a+\lambda b)^2 - (a-\lambda b)^2\;, \eqno{(3.17)}$$\\[-4ex]

where $\lambda$ is a power of 2 chosen so that

$$\textstyle \frac{1}{2} \leq |\lambda b/a| \leq 2 
                                 \eqno{(3.18)}$$\\[-4ex]

(unless $a=0$ or $b=0)$.  This scaling is necessary to avoid excessive
cancellation in (3.17).  (A~detailed discussion of a similar situation
is given in Brent $[5]$.)  From (3.17), 

$$ t_n(M) \leq 2t_n(S) + 3t_n(A) + c_9, \eqno{(3.19)}$$\\[-4ex]

so $M \less S$ follows from Lemma 3.1.\\

The next two lemmas show that multiple-precision multiplication
is linearly equivalent to taking reciprocals and to division.  The
idea of the proof of Lemma 3.6 is to use a Newton iteration involving
only multiplications and additions to approximate $1/a$.
Computational work is saved by starting with low precision and
approximately doubling the precision at each iteration.  The basic idea
is well-known and has even been implemented in hardware.\\

The possibility of saving work by increasing the precision at each 
iteration is examined more closely in Sections 7 and 8.\\

\begin{lemma}
$$I \less D \less M\;. \eqno{(3.20)}$$\\[-4ex]
\end{lemma}

{\bf Proof.} Consider the iteration

$$ x_{j+1} = x_j\left(2-ax_j\right) \eqno{(3.21)}$$\\[-4ex]

obtained by applying Newton's method to the equation $x^{-1}-a=0$.
If

$$ x_j=\left(1-\varepsilon_j\right)a^{-1}\;, \eqno{(3.22)}$$\\[-4ex]

then substitution in (3.21) shows that

$$ \varepsilon_{j+1}=\varepsilon^2_j\;, \eqno{(3.23)}$$\\[-4ex]

so the order of convergence is two.  A single-precision computation
is sufficient to give an initial approximation such that
$|\varepsilon_0| \leq \frac{1}{2}$, and it follows from
(3.23) that

$$|\varepsilon_j| \leq 2^{-2^j} \eqno{(3.24)}$$\\[-4ex]

for all $j \geq 0$.\\

In deriving (3.24) we have assumed that (3.21) is satisfied exactly,
but a result like (3.24) holds so long as the right hand side of
(3.21) is evaluated using a precision of at least $2^{j+1}$ bits.
Thus, an $n$-bit approximation to $a^{-1}$ can be obtained by
performing $\lceil \log_2n\rceil$ iterations of (3.22) with precision
at least $2, 2^2, 2^3, \ldots, 2^{\lceil\log_2n\rceil -1}$, $n$
at each iteration.  From Lemma 3.4 (with $\alpha = \frac{1}{2}$),
this gives

$$ t_n(I) \leq c_{10}t_n(M)\;. \eqno{(3.25)}$$\\[-4ex]

Since $b/a = b(1/a)$, it follows that

$$ t_n(D) \leq c_{11}t_n(M)\;, \eqno{(3.26)}$$\\[-4ex]

so $D\less M$.  Since $I \less D$ is trivial, the proof is complete.\\

{From} $ab=a/(1/b)$ it is clear that $M\less D$.  The proof that
$M \less I$ is not quite so obvious, and uses the equivalences of
multiplication and squaring (Lemma 3.5).\\

\begin{lemma}
$$ M \less I\;. \eqno{(3.27)}$$\\[-4ex]
\end{lemma}

{\bf Proof.}  We may apply the identity

$$ a^2(1-\lambda a)^{-1} = \lambda^{-2}\left[(1-\lambda a)^{-1} - 
(1+\lambda a)\right] \eqno{(3.28)}$$\\[-4ex]

to obtain an approximation to $a^2$, using only the operation of
taking reciprocals, addition (or subtraction) and multiplication by
powers of two.  If $a \neq 0$, choose $\lambda$ to be a power of two
such that

$$ 2^{-n/3-1} < |\lambda a| < 2^{1-n/3}\;, 
                                          \eqno{(3.29)}$$\\[-4ex]

and evaluate the right hand side of (3.28), using precision $n$.  
This gives an approximation to $a^2$ with precision $\lceil n/3 \rceil$,
so

$$ S_{\lceil n/3\rceil} \less I_n\;, \eqno{(3.30)}$$\\[-4ex]

where the subscripts denote the precision.  Thus, the result follows
from Lemmas 3.2 and 3.5.\\

To conclude this section we consider the complexity of multiple-precision
square roots.  Results like Lemmas 3.8 and 3.9 actually hold if 
$x^{\frac{1}{2}}$ is replaced by $x^p$ for any fixed rational
$p \neq 0$ or $1$ (we have already shown this for $p=-1$).\\

\begin{lemma}
$$M\less R\;. \eqno{(3.31)}$$\\[-4ex]
\end{lemma}

{\bf Proof.}  The proof is similar to that of Lemma 3.7, using the
approximation\\ $2\lambda^{-2}\left[1+\lambda a - 
(1+2\lambda a)^{\frac{1}{2}}\right]$ to $a^2$.\\

\begin{lemma}
$$ R \less M\;. \eqno{(3.32)}$$\\[-4ex]
\end{lemma}

{\bf Proof.}  The proof is similar to that of Lemma 3.6, using
Newton's iteration

$$ x_{j+1} = {\textstyle\frac{1}{2}}\left(x_j+a/x_j\right)\;, 
                                             \eqno{(3.33)}$$\\[-4ex]

with precision increasing at each iteration, to approximate 
$\sqrt{a}$.
Alternatively, it is possible to avoid multiple-precision division
by using the iteration
\pagebreak

$$x_{j+1} = x_j\left(3-ax^2_j\right)/2 \eqno{(3.34)} $$\\[-4ex]

to approximate $a^{-\frac{1}{2}}$, and then use 
$ \sqrt{a} = a.a^{-\frac{1}{2}}$ to evaluate $\sqrt{a}$.\\

The results of Lemmas 3.5 to 3.9 may be summarized in the following:\\

\begin{theorem}
$$ D \equiv I \equiv M \equiv R \equiv S\;. \eqno{(3.35)}$$\\[-4ex]
\end{theorem}


\section{Some regularity conditions}

Before discussing the complexity of multiple-precision evaluation
of exponentials, trigonometric functions, etc., we need some definitions.
Throughout this section, let $\phi(x)$ be a real-valued function which 
is positive and monotonic increasing for all sufficiently large positive
$x$.\\

\begin{definition} \rm
$\phi \in \Phi_1$ iff, for all $\alpha \in (0,1)$, for some positive
$K$, for all sufficiently large $x$ and all $x_0, \ldots, x_J$
satisfying

$$ 1 \leq x_j \leq \alpha^jx \eqno{(4.1)}$$\\[-4ex]

for $j=0, \ldots, J$, we have

$${\sum^J_{j=0}} \phi(x_j) \leq K\phi(x)\;. \eqno{(4.2)}$$\\[-2ex]

$\phi \in \Phi_2$ iff, for some $\alpha, \beta \in (0,1)$ and all
sufficiently large $x$,

$$\phi(\alpha x) \leq \beta\phi(x)\;. \eqno{(4.3)}$$\\[-4ex]

$\phi \in \Phi_3$ iff, for some positive $K_1, K_2$ and $p$, there
is a monotonic increasing function $\psi$ such that

$$K_1x^p\psi(x) \leq \phi(x) \leq K_2x^p\psi(x) \eqno{(4.4)}$$\\[-4ex]

for all sufficiently large $x$.
\end{definition}

Note the similarity between the definition of $\Phi_1$ and the statement
of Lemma 3.4.  In Section 5, we need to assume that the time $\phi(n)$
required to perform certain operations with precision $n$ satisfies
(4.2).  The following lemmas make this assumption highly plausible.
Lemma 4.1 shows that ``for all $\alpha$'' in the definition of $\Phi_1$
may be replaced by ``for some $\alpha$''.\\

\begin{lemma}
If, for some $\alpha \in (0,1)$ and some positive $K$, for all sufficiently
large $x$ and all $x_0, \ldots, x_J$ satisfying {\rm{(4.1)}}, 
we have {\rm{(4.2)}},
then $\phi \in \Phi_1$.
\end{lemma}

{\bf Proof.}  Take any $\alpha_1$ and $\alpha_2$ in (0, 1), and suppose
that (4.1) with $\alpha$ replaced by $\alpha_2$ implies (4.2) with $K$
replaced by $K_2$.  Let $m$ be a positive integer such that 
$\alpha^m_1 \leq \alpha_2$.  If (4.1) holds with $\alpha$ replaced by
$\alpha_1$ for a sequence $(x_0, x_1, \ldots, x_J)$, then (4.1) also
holds with $\alpha$ replaced by $\alpha_2$ for each of the $m$
subsequences

$$(x_0, x_m, \ldots), (x_1, x_{m+1}, \ldots), \ldots, (x_{m-1}, x_{2m-1}, 
         \ldots)\;,$$\\[-4ex]

so (4.2) holds with $K$ replaced by $K_1 = mK_2$.\\

Lemmas 4.2 and 4.3 show that $\phi \in \Phi_2$ or $\phi \in \Phi_3$
is a sufficient condition
for $\phi \in \Phi_1$.  The proof of Lemma 4.2 is similar to that of
Lemma 3.4 (using Lemma 4.1), so is omitted.\\

\begin{lemma}
$$ \Phi_2 \subseteq \Phi_1\;. \eqno{(4.5)}$$\\[-4ex]
\end{lemma}
\vspace*{-2mm}
\begin{lemma}
$$ \Phi_2 = \Phi_3\;. \eqno{(4.6)}$$\\[-4ex]
\end{lemma}

{\bf Proof.}  First suppose that $\phi \in \Phi_3$, so (4.4) holds for
some function $\psi$ and some positive $K_1, K_2$ and $p$.  Choose
$\alpha \in (0,1)$ such that $\beta = \alpha^p K_2/K_1 < 1$.  For all
sufficiently large $x$, we have

$$ \phi(\alpha x) \leq K_2\alpha^px^p\psi(\alpha x) 
\leq K_2\alpha^px^p\psi(x) \leq \left(K_2\alpha^p/K_1\right)\phi(x)
\leq \beta\phi(x)  \eqno{(4.7)}$$\\[-4ex]

(using (4.4) and the monotonicity of $\psi)$, so $\phi \in \Phi_2$.\\

Conversely, suppose that $\phi \in \Phi_2$, so (4.3) holds for all
sufficiently large $x$ (say $x \geq x_0>0$) and some 
$\alpha, \beta \in (0,1)$. Choose $p$ small enough that 
$\beta \leq \alpha^p$, so 

$$\phi(\alpha x) \leq \alpha^p\phi(x) \eqno{(4.8)}$$\\[-4ex]

for $x \geq x_0$. Since $\phi(x)$ is positive for sufficiently large $x$,
we may assume that $\phi(x_0) > 0$. Let $K_1 = \alpha^p, K_2 = 1$, and

$$ \psi(x) = \sup_{x_0\leq y\leq x} \phi(y)/y^p \eqno{(4.9)}$$\\[-2ex]

for $x \geq x_0$.  Thus, $\psi(x)$ is monotonic increasing and 

$$ \psi(x) \geq \phi(x)/x^p \eqno{(4.10)}$$\\[-4ex]

so

$$ \phi(x) \leq K_2x^p\psi(x) \eqno{(4.11)}$$\\[-4ex]

for $x \geq x_0$.\\

By repeated application of (4.8) we have, for $k \geq 0$,

$$\phi(x)/x^p \geq \phi(\alpha^kx)/(\alpha^kx)^p \eqno{(4.12)}$$\\[-4ex]

provided $\alpha^kx \geq x_0$.  Thus, from (4.9),

$$ \psi(x) = \sup_{\alpha x\leq y\leq x} \phi(y)/y^p \eqno{(4.13)}$$\\[-4ex]

$$ \;\;\; \leq \phi(x)/(\alpha x)^p \eqno{(4.14)}$$\\[-4ex]

for $x\geq x_0/\alpha$.  Thus,

$$ \phi(x) \geq K_1x^p\psi(x) \eqno{(4.15)}$$\\[-4ex]

and, in view of (4.11), $\phi \in \Phi_2$. 


\section{Linear equivalence of various elementary functions}

In this section, we consider the multiple-precision ``operations''
of evaluating certain elementary functions (log, exp, sin, artan,
etc).  First we prove three theorems which apply under fairly 
general conditions.  Theorem 5.1 is a generalization of Lemmas 3.7
and 3.8, and gives a simple condition under which the evaluation of
$f(x)$ is at least as difficult as a multiplication (in the sense
of Definition 1.1).\\

{\bf NOTATION.}  If $f$ is a real-valued function defined on some 
finite interval $[a, b]$, the operation of evaluating $f(x)$ to
(relative) precision $n$ for $x\in[a,b]$ is denoted by 
$E^{(n)}_{[a,b]}(f)$.  If there is no risk of confusion, we write
simply $E_{[a,b]}(f)$ or $E(f)$.  We sometimes write $t_n(f)$ for
$t_n\left(E(f)\right)$.  $LC^{(m)}[a,b]$ is the class of functions
with Lipschitz continuous $m$-th derivatives on $[a,b]$.  We always
assume that $b>a$.\\

\begin{theorem}
If $f\in LC^{(2)}[a,b]$ and there is a point $x_0\in (a,b)$ such
that $f''(x_0) \neq 0$, then

$$ E(f) \geq M\;.  \eqno{(5.1)}$$\\[-4ex]
\end{theorem}

{\bf Proof.}  For all sufficiently small $h$, we have (from 
$[$2, Lemma 3.2$]$)

$$ f(x_0+h) + f(x_0-h) - 2f(x_0) = h^2f''(x_0) + R(x)\;, \eqno{(5.2)}$$\\[-4ex]

where

$$|R(x)| \leq c_{12}|h|^3\;. 
                                    \eqno{(5.3)}$$\\[-4ex]

Let $c = f''(x_0) \neq 0$.  Three evaluations of $f$ and some additions
may be used to approximate $ch^2$, using (5.2).  If $h$ is of  order
$2^{-n/3}$, the resulting approximation to $ch^2$ has relative error of
order $2^{-n/3}$.  Proceeding as in the proof of Lemma 3.5, we see that
six evaluations of $f$ and some additions may be used to approximate
$cxy$ to precision $n/3$, for any $x$ and $y$.  Applying this result,
with $x$ replaced by the stored constant $c^{-2}$, and $y$ replaced by
the computed $cxy$, shows that 12 evaluations of $f$ give 
$c\left(c^{-2}\right)(cxy) = xy$ to precision $n/3$.  The result now 
follows from Lemma 3.2. \\

{\bf REMARK.}  If $f''(x)$ is not constant on $[a,b]$, the point
$x_0$ may be chosen so that $f''(x)$ is rational, so (5.2) may be
used to approximate $h^2$, and the result follows more easily (as
in the proof of Lemma 3.7).\\

Theorem 5.2 gives conditions under which the multiple-precision
evaluation of the inverse function $g=f^{(-1)}$ of a function $f$
is linearly reducible to the evaluation of $f$.  (The inverse function
satisfies $g\left(f(x)\right)=x$.)  The condition $0\not\in [a,b]$ could
be dropped, if we only required the computation of $g$ with an absolute
(rather than relative) error of order $2^{-n}$.\\

\begin{theorem}
If $0 \not\in [a,b]$, $f \in LC^{(1)}[a,b]$, $f^{\prime}(x) \neq 0$
on $[a,b]$, $E(f) \more M$, and
$$ t_n(f) \in \Phi_1\;, \eqno{(5.4)} $$\\[-4ex]

then

$$ E(g) \less E(f)\;, \eqno{(5.5)}$$\\[-4ex]

where $g = f^{(-1)}$ and $\Phi_1$ is as in Definition 4.1.
\end{theorem}

{\bf Proof.}  Since $f'(x)$ is continuous and nonzero on $[a, b]$,
there is no loss of generality in assuming that

$$ f'(x) \geq c_{13} > 0 \eqno{(5.6)} $$\\[-4ex]

on $[a, b]$.  Thus, $g(y)$ exists on $[c, d] = [f(a), f(b)]$.
Also, since $0 \neq [a, b]$, we have

$$|g(y)| \geq c_{14} > 0 \eqno{(5.7)}$$\\[-4ex]

on $[c, d]$.\\

To estimate $g(y)$ we may solve $\psi(x)=0$ by a discrete version
of Newton's method, where 

$$ \psi(x) = f(x) - y \;. \eqno{(5.8)}$$\\[-4ex]

Consider the iteration

$$ x_{j+1} = x_j - \psi(x_j)/\mu_j\;, \eqno{(5.9)}$$\\[-4ex]

where

$$ \mu_j = \left(\psi(x_j+h_j)-\psi(x_j)\right)/h_j\;, 
                                         \eqno{(5.10)}$$\\[-4ex]

and the computation of $\mu_j$ and $x_{j+1}$ is performed with precision
$n_j \leq n$, giving computed values $\widehat{\mu}_j$ and $\widehat{x}_{j+1}$
respectively.  If $h_j$ is of order $2^{{-n_j}/2}$, then

$$|\widehat{\mu}_j - \psi'(\widehat{x}{_j})| \leq\; 2^{-n_j/2}c_{15}\;,
                                    \eqno{(5.11)}$$\\[-4ex]

and it is easy to show that

$$|\widehat{x}_{j+1}-g(y)| \;\leq\; 
|\widehat{x}_j-g(y)|^2 \; c_{16}
\; + \; 2^{-n_j /2}\; |\widehat{x}_j - g(y)| \; c_{17} 
\; + \; 2^{-n_{j}} \; c_{18}\;.
                                 \eqno{(5.12)}$$\\[-4ex]

Since a sufficiently good starting approximation $x_0$ may be found
using single-precision (or at most bounded-precision) computation, 
(5.12) ensures that

$$|\widehat{x}{_{j+1}}-g(y)| \;\leq\; 
|\widehat{x}{_j}-g(y)|^2\; c_{19}\;,
                                \eqno{(5.13)}$$\\[-4ex]

provided

$$|\widehat{x}_j -g(y)| \;\geq\; 2^{-n_j/2}\;. \eqno{(5.14)}$$\\[-4ex]

Hence, we may approximately double the precision at each iteration, and
(5.13) guarantees convergence of order two.  A final iteration with
$h_j = 2^{-n/2}$ will be sufficient to give

$$|\widehat{x}_{j+1}-g(y)| \;\leq\; 2^{-n}c_{20}\;.
                                    \eqno{(5.15)}$$\\[-4ex]

Since $E(f) \geq M$, the result follows from (5.4), (5.7), (5.15),
and Lemma 3.6.\\

\begin{theorem}
If $0 \not\in [a, b]$, $f \in LC^{(1)}[a, b]$, $f(x)f'(x) \neq 0$
on $[a, b]$, $g=f^{(-1)}$, $E(f) \more M$, $E(g) \more M$,
$t_n(f) \in \Phi_1$, and $t_n(g) \in \Phi_1$, then

$$ E(f) \equiv E(g)\;. \eqno{(5.16)}$$\\[-4ex]
\end{theorem}

{\bf Proof.}  Since $t_n(f) \in \Phi_1$, Theorem 5.2 applied to $f$
gives $E(g) \less E(f)$.  Similarly, applying Theorem 5.2 to $f^{(-1)}$
gives $E(f) \less E(g)$, so the result follows.\\

We are now ready to deduce the linear equivalence of $mp$ evaluation
of various elementary functions $f_i$, assuming that $t_n(f_i) \in
\Phi_1$. In view of Lemmas 4.2 and 4.3, this assumption is very 
plausible.\\

\begin{corollary}
If $0<a<b$, $c<d$, $1 \not\in [a, b]$, $t_n\left(E_{[a,b]}(\log)\right)
\in \Phi_1$, and $t_n\left(E_{[c,d]}(\exp)\right) \in \Phi_1$, then

$$ E_{[a,b]}(\log) \equiv E_{[c,d]}(\exp)\;. \eqno{(5.17)}$$\\[-4ex]
\end{corollary}

{\bf Proof.}  From Theorem 5.1, $E_{[a,b]}(\log)\more M$ and
$E _{[c,d]}(\exp)\more M$.  Also, the identities

$$ \exp(-x) = 1/\exp(x) \eqno{(5.18)}$$\\[-6ex]

and

$$ \exp(\lambda x) = \left(\exp(x)\right)^{\lambda} \eqno{(5.19)}$$\\[-4ex]

(for suitable rational $\lambda$) may be used to show that
$E_{[c,d]}(\exp) \equiv E_{[c',d']}(\exp)$ for any $c' < d'$.
Hence, the result follows from Theorem 5.3.\\

{\bf REMARK.}  
If $1 \in [a, b]$, then Theorem 5.2 shows that

$$ E^{(n)}_{[c,d]}(\exp) \less E^{(n)}_{[a,b]}(\log)\;, \eqno{(5.20)}$$\\[-4ex]

and a proof like that of Theorem 5.2 shows that

$$ E^{(n)}_{[a,b]}(\log) \less E^{(2n)}_{[c,d]}(\exp)\;, \eqno{(5.21)}$$\\[-4ex]

so the conclusion of Corollary 5.1 follows, if

$$E^{(2n)}_{[c,d]}(\exp) \equiv E^{(n)}_{[c,d]}(\exp)\;. \eqno{(5.22)}$$\\[-4ex]

Although (5.22) is plausible, no proof of it is known.  (The corresponding
result for multiplication is given in Lemma 3.2.)\\

\begin{corollary}

\begin{eqnarray*}
E(\sinh) &\equiv& E(\cosh) \equiv E(\tanh) \equiv E(\arsinh)\\ 
	 &\equiv& E(\arcosh) \equiv E(\artanh) \equiv E(\exp) \equiv E(\log)
\end{eqnarray*}\\[-45pt]$$ \eqno{(5.23)} $$ 

on any nontrivial closed intervals on which the respective functions
are bounded and nonzero, assuming $t_n(\sinh) \in \Phi_1$ etc.
\end{corollary}  \vspace*{3mm}

\begin{corollary} 

$$E(\sin) \equiv E(\cos) \equiv E(\tan) \equiv E(\arsin) \equiv E(\arcos)
\equiv E(\artan) \eqno{(5.24)}$$

on any nontrivial closed intervals on which the respective functions
are bounded and nonzero, assuming $t_n(\sin) \in \Phi_1$ etc.
\end{corollary} 
\vspace*{3mm}
{\bf REMARKS.}  The proofs of Corollaries 5.2 and 5.3 are similar to
that of Corollary 5.1 (using well-known identities), so are omitted.
Since $\exp(ix) = \cos(x) + i \sin(x)$, it is plausible that
$E(\exp) \equiv E(\sin)$, but we have not proved this.  (It is just 
conceivable that the evaluation of $\exp(x)$ for complex $x$ is not
linearly reducible to the evaluation of $\exp(x)$ for real $x$.)


\section{Upper and lower bounds}

In this section we give some upper and lower bounds on $t_n(A)$,
$t_n(M)$, $t_n(\exp)$ and $t_n(\sin)$.  Since the multiplicative 
constants are not specified, the bounds apply equally well to the
operations which are linearly equivalent to addition, multiplication,
etc.~(see Sections 2 to 5).  The lower bounds are trivial:
$t_n{\exp\choose\sin} \geq c_{21}t_n(M) \geq c_{22}t_n(A)
\geq c_{23}n$ (from (2.2), Lemma 3.1 and Theorem 5.1). The upper
bounds are more interesting.  \\

{\bf UPPER BOUNDS ON $t_{n}(M)$}\\

The obvious algorithm for multiplication of multiple-precision 
numbers gives

$$ t_n(M) \leq c_{24}n^2\;, \eqno{(6.1)}$$\\[-4ex]

but this is not the best possible upper bound.  Karatsuba and Ofman
$[12]$ showed that

$$ t_n(M) \leq c_{25}n^{1.58\ldots}\;, \eqno{(6.2)}$$\\[-4ex]

where $1.58 \ldots \;=\; \log_{2}3$.  The idea of the proof is that,
to compute

$$ (a+\lambda b)(c+\lambda d)= ac+ \lambda(ad+bc) + \lambda^2bd\;,
                                 \eqno{(6.3)}$$\\[-4ex]

where $\lambda$ is a suitable power of two, we compute the three
products $m_1=ac$, $m_2=bd$, and $m_3=(a+b)(c+d)$, and use the
identity

$$ ad+bc=m_3 - (m_1+m_2)\;. \eqno{(6.4)}$$\\[-4ex]

Thus, $2n$-bit integers can be multiplied with three multiplications 
of (at most) $(n+1)$-bit integers, some multiplications by powers of 
two, and six additions of (at most) $4n$-bit integers.  This 
observation leads to a recurrence relation from which (6.2) follows.\\

More complicated identities like (6.4) may be used to reduce the
exponent in (6.2).  Recently Sch\"{o}nhage and Strassen $[23]$
showed that the exponent can be taken arbitrarily close to unity.
Their method gives the best known upper bound

$$ t_n(M) \leq c_{26}n\;  \log(n)\log\log(n)\;, \eqno{(6.5)}$$\\[-4ex]

and uses an algorithm related to the fast Fourier transform to compute
certain convolutions.  For a description of this and earlier methods
see Knuth $[13$ (revised)$]$.  Knuth conjectures that (6.5) is
optimal, though the term $\log \log(n)$ is rather dubious.  (It may
be omitted if a machine with random-access memory of size $O(n^p)$
for some fixed positive $p$ is assumed.)  From results of
Morgenstern $[19]$ and Cook and Aanderaa $[8]$, it is extremely
probable that

$$ \lim_{n\rightarrow\infty} t_n(M)/n = \infty\;, \eqno{(6.6)}$$\\[-4ex]

which implies that $M \not\equiv A$, but more work remains to be
done to establish this rigorously.\\

{\bf UPPER BOUNDS ON} $t_n(\exp)$ {\bf AND} $t_n(\sin)$\\

To evaluate $\exp(x)$ to precision $n$ from the power series

$$ \exp(\pm x) = \sum^{\infty}_{j=0} (\pm x)^j /j!\;, \eqno{(6.7)}$$\\[-4ex]

it is sufficient to take $c_{27}n/\log(n)$ terms, so

$$ t_n(\exp) \leq c_{28}t_n(M)n/\log(n)\;. \eqno{(6.8)}$$\\[-4ex]

Theorem 6.1 shows that the bound (6.8) may be reduced by a factor of
order $\sqrt{n}/\log(n)$.\\

\begin{theorem}
$$ t_n(\exp) \leq c_{29} \sqrt{n} \; t_n(M) \eqno{(6.9)}$$\\[-4ex]

and

$$ t_n(\sin) \leq c_{30} \sqrt{n} \; t_n(M)\;. \eqno{(6.10)}$$\\[-4ex]
\end{theorem}

{\bf Proof.} To establish (6.9), we use the identity

$$ \exp(x) = \left(\exp(x/\lambda)\right)^\lambda  \eqno{(6.11)}$$\\[-4ex]

with $\lambda = 2^q$, where $q=\lfloor n^{\frac{1}{2}}\rfloor$.
If $[a, b]$ is the domain of $x$, 
and $c = \max(|a|, |b|)$, then

$$|(x/\lambda)^r/r!| \;\leq\; 2^{-qr}\;, \eqno{(6.12)}$$\\[-4ex]

if $r$ is large enough that

$$ c^r \leq r!\;. \eqno{(6.13)}$$\\[-4ex]

Hence, it is sufficient to take $r = \lceil n/q \rceil$ terms in the
power series for $\exp(x/\lambda)$ to give an absolute error of order 
$2^{-n}$ in the approximation to $\exp(x/\lambda)$.  Since
$\exp(x/\lambda)$ is close to unity, the relative error will also
be of the order $2^{-n}$ for large $n$.  From (6.11), $q$ squarings
may be used to compute $\exp(x)$ once $\exp(x/\lambda)$ is known.\\

The method just described gives $\exp(x)$ 
to precision $n - n^{\frac{1}{2}}$, for the relative error in
$\exp(x/\lambda)$ is amplified by the factor $\lambda$.  This may
be avoided by taking $r = \lceil n/q \rceil +1$, and either working
with precision $n + n^{\frac{1}{2}}$, or evaluating

$$ \exp(|x/\lambda|) - 1 \simeq \sum^{r}_{j=1}
|x/\lambda|^j/j! \eqno{(6.14)}$$\\[-4ex]

and then using the identity

$$ (1+\varepsilon)^2 - 1 = 2\varepsilon + \varepsilon^2 
                                       \eqno{(6.15)}$$\\[-4ex]

to evaluate $\exp(|x|) - 1$ without appreciable loss
of significant figures.  Thus, (6.9) follows (using Lemma 3.2 if 
necessary).\\

The proof of (6.10) is similar, using the identity

$$ \sin(x) = \pm2 \sin(x/2) \sqrt{1-\sin^2(x/2)}\;,
                              \eqno{(6.16)}$$\\[-4ex]

$q$ times to reduce the computation of $\sin(x)$ to that of
$\sin(x/\lambda)$ (recall Lemma 3.9).\\

{\bf REMARKS}.  If $x$ is a rational number with small numerator and
denominator, the time required to sum $r$ terms in the power series
for $\exp(x/\lambda)$ is $O(rn)$, and the time required for $q$
squarings is $O\left(qt_n(M)\right)$.  Thus, choosing
$r = \lfloor \sqrt{t_n(M)} \rfloor$ and
$q = \lceil n/r \rceil$ gives total time 
$O \left( n \sqrt{t_n(M)}\right)$.
It is also possible to evaluate $\exp(x)$ in this time for general 
$x$, by using a form of preconditioning to reduce the number of
multiplications required to evaluate the power series for
$\exp(x/\lambda)$.~\\~\\

{\bf A NUMERICAL EXAMPLE}\\

The following example illustrates the ideas of Theorem 6.1.  Suppose 
we wish  to calculate $e$ to 30 decimal places.  The obvious method is 
to use the approximation

$$ e \simeq \sum^{28}_{j=0} 1/j! \eqno{(6.17)}$$\\[-4ex]

(since $29! \simeq 8.8 \times 10^{30}$).  On the other hand

$$ e \simeq \left(\sum^{10}_{j=0} \frac{1}{j! \; 256^j}\right)^{256}
                                   \eqno{(6.18)}$$\\[-2ex]

also gives the desired accuracy 
(since $11! \; 256^{10} \simeq 4.8 \times 10^{31}$).  
Thus, the computation of 18 inverse factorials may be 
saved at the expense of 8 squarings.\\

Similarly, the computation of $e$ to $10^6$ decimal places by the obvious 
method requires the sum of about 205,030 inverse factorials, but the
approximation

$$ e \simeq \left( \sum^{1819}_{j=0} 
	\frac{1}{j! \; 2^{1820j}}\right)^{2^{1820}}\;,
                         \eqno{(6.19)}$$\\[-2ex]

requiring only 1820 terms and 1820 squarings, is sufficiently accurate.\\

{\bf BASE CONVERSION}\\

Sch\"{o}nhage has shown that conversion from binary to decimal or vice 
versa may be done in time 
$O\left(n\left(\log(n)\right)^2\log\left(\log(n)\right)\right)$
(see Knuth $[13, {\rm ex.}\; 4.4.14\; ({\rm revised})]$).  We describe his
method here, as a similar idea is used below to improve Theorem 6.1.\\

Let $\beta > 1$ be a fixed base (e.g.~$\beta = 10$), and suppose we know 
the base $\beta$ representation of an integer $x$, i.e.~we know the
digits $d_0, \ldots, d_{t-1}$, where $0 \leq d_i < \beta$ and
$\displaystyle x = \sum^{t-1}_{0} d_i \beta^i$.  Suppose that $n$-bit  binary
numbers can be multiplied exactly in time $M(n)$, where

$$ 2M(n) \leq M(2n) \eqno{(6.20)}$$\\[-4ex]

for all sufficiently large $n$.  (This is certainly true if the 
Sch\"{o}nhage-Strassen method $[13, 23]$ is used.)  We describe how 
the binary representation of $x$ may be found in time
$O\left(M(n)\log(n)\right)$, where $n$ is sufficiently large for $x$ to be 
representable as an $n$-bit number (i.e.~$2^n \geq \beta^t$).\\

Without changing the result, we may suppose $t = 2^k$ for some
positive integer $k$.  Let the time for conversion to binary and
computation of $\beta^{2^k}$ be $C(k)$.  Thus, we can compute
$\beta^{t/2}$ and convert the numbers
$\displaystyle x_1 = \sum^{t/2-1}_{0}d_i\beta^i$ and 
$\displaystyle x_2 = \sum^{t-1}_{t/2} d_i \beta^{i-t/2}$ to binary in time
$2C(k-1)$, and then $x=x_1+\beta^{t/2}x_2$ and 
$\beta^t=\left(\beta^{t/2}\right)^2$ may be computed in time
$2M(n/2)+O(n)$.  Thus

$$ C(k) \leq 2C(k-1)+2M(n/2)+O(n)\;, \eqno{(6.21)}$$\\[-4ex]

so

$$ C(k) \leq 2M(n/2)+ 4M(n/4) + 8M(n/8) + \ldots + O(n\log(n))$$

$$ \hspace*{-54mm} \leq O(M(n)\log(n)) \eqno{(6.22)}$$\\[-4ex]

(using (6.20)).\\

The proof that conversion from base 3 to base $\beta$ may be done
in time (6.22) is similar, and once we can convert integers it is
easy to convert floating-point numbers.\\

{\bf COMPUTATION OF} $e$ {\bf AND} $\pi$\\

We may regard $e-2 = 1/2! + 1/3! + \ldots\;$ as given by a mixed-base
fraction $0.111 \ldots$, where the base is $2, 3, \ldots$  Hence,
it is possible to evaluate $e$ to precision $n$, using a slight
modification of the above base-conversion method, in time
$O(M(n)\log(n))$.\\

Similarly, $\artan(1/j)$ may be computed to precision $n$ in time
$O(M(n)\log^2(n))$, for any small integer $j \geq 2$, and then
$\pi$ may be computed from well-known identities such as Machin's

$$ \pi = 16 \; \artan(1/5) - 4 \; \artan(1/239)\;. \eqno{(6.23)}$$\\[-4ex]

The methods just described are asymptotically faster than the $O(n^2)$
methods  customarily used in multiple-precision calculations of $e$
and $\pi$ (see, for example, Shanks and Wrench $[25, 26]$).  It
would be interesting to know how large $n$ has to be before the
asymptotically faster methods are actually faster.  A proof that even
faster methods are impossible would be of great interest, for it would
imply the transcendence of $e$ and $\pi$.\\

{\bf IMPROVED UPPER BOUNDS ON} $t_n(\exp)$ {\bf AND} $t_n(\sin)$\\

The following lemma uses an idea similar to that described above for
base conversion and computation of $e$.\\

\begin{lemma}
If $p$ and $q$ are positive integers such that $p^2 \leq q \leq 2^n$,
then $\exp(p/q)$ may be computed to precision $n$ in time 
$O(M(n)\log(n))$.
\end{lemma}

{\bf Proof.}  The approximation

$$ \exp(p/q) \simeq \sum^{k}_{j=0} \frac{(p/q)^j}{j!} \eqno{(6.24)}$$\\[-2ex]

is sufficiently accurate if $k$  is chosen so that

$$\frac{(p/q)^{k+1}}{(k+1)!} \leq 2^{-n} \leq \frac{(p/q)^k}{k!}\;.
                             \eqno{(6.25)}$$\\[-2ex]

Since $p^2 \leq q$, (6.25) gives $k!q^{k/2} \leq 2^n$, so certainly

$$ k!q^k \leq 2^{2n}\;, \eqno{(6.26)} $$\\[-4ex]

Hence, a method like that described above for the computation of
$e$ may be used, and (6.26) ensures that the integers in intermediate 
computations do not grow too fast.\\

{From} Lemma 6.1 it is easy to deduce Theorem 6.2, which is an
improvement of Theorem 6.1 for large $n$.  The methods used in the
proof of Theorem 6.1 and the following remarks are, however, faster
than that of Theorem 6.2 for small and moderate values of $n$.\\


\begin{theorem}
If $M(n)$ satisfies {\rm (6.20)} then

$$ t_n(\exp) \leq c_{32}M(n)\log^2(n) \eqno{(6.27)}$$\\[-4ex]

and

$$ t_n(\sin) \leq c_{33}M(n)\log^2(n)\;.  \eqno{(6.28)}$$\\[-4ex]
\end{theorem}

{\bf Proof.}  Without affecting the result (6.27) we may assume
that $n = 2^k$ for some positive integer $k$.  (This assumption 
simplifies the proof, but it is not essential.)  Given an $n$-bit
fraction $x \in [0, 1)$, we write

$$ x = \sum^{k}_{i=0} p_i/q_i\;, \eqno{(6.29)}$$\\[-2ex]

where $q_i = 2^{2^i}$ and $0 \leq p_i < 2^{2^{i-1}}$ for 
$i = 0, 1, \ldots, k$.  By Lemma 6.1, $\exp(p_i/q_i)$ can be computed,
to sufficient precision, in time $O(M(n)\log(n))$, so

$$\exp(x) = \prod^{k}_{i=0} \exp(p_i/q_i) \eqno{(6.30)}$$\\[-2ex]

can be computed in time $O(M(n)(\log(n))^2)$.  This establishes
(6.27), and the proof of (6.28) is similar.\\

\begin{corollary}

$$ t_n(\exp) \leq c_{34}n(\log(n))^3\log\log(n) \eqno{(6.31)}$$\\[-4ex]

and

$$ t_n(\sin) \leq c_{35}n(\log(n))^3\log\log(n)\;.  \eqno{(6.32)}$$\\[-4ex]
\end{corollary}

{\bf Proof.} This is immediate from the bound (6.5) and Theorem 6.2.\\

\begin{corollary}

$$ t_n(E_{[a,b]}(f)) \leq c_{36}n(\log(n))^3\log\log(n)\;, $$\\[-4ex]

where
\begin{eqnarray*}
f(x) &=& \log(x), \exp(x), \sin(x), \cos(x), \tan(x), \sinh(x),\\
     & & \cosh(x), \tanh(x), \arsin(x), \artan(x), \arsinh(x),
\end{eqnarray*}
etc, and $[a, b]$ is any finite interval on which $f(x)$ is bounded.
\end{corollary}

{\bf Proof.}  This follows from (6.5), Corollaries 5.1 (and the note
following), 5.2, 6.1, and Lemma~3.2.


\section{Best constants for operations equivalent to multiplication}

In this section, we consider in more detail the relationship between
the $mp$ operations $D, I, M, R$, and $S$ defined in Section 3.  It
is convenient to consider also the operation $Q$ of forming inverse
square roots (i.e., $y \leftarrow x^{-\frac{1}{2}}$).  From Theorem 3.1,
if we can perform any one of these operations (say $Y$) to precision
$n$ in time $t_n(Y)$, then the time required to perform any of the other
operations to precision $n$ is at most a constant multiple of
$t_n(Y)$.\\

\begin{definition} \rm
$C_{XY}$ is the minimal constant such that, for all positive $\varepsilon$
and all sufficiently large $n$, the operation $X$ can be performed (to
precision $n$) in time $(C_{XY}+\varepsilon)t_n(Y)$ if $Y$ can be
performed in time $t_n(Y)$, where $X, Y=D, I, M, Q, R$ or $S$.
\end{definition}

The following inequalities are immediate consequences of Definition 7.1:

$$ C_{XY}C_{YZ} \geq C_{XZ} \eqno{(7.1)}$$\\[-4ex]
\pagebreak

and

$$ C_{XY}C_{YX} \geq C_{XX} = 1\;. \eqno{(7.2)}$$\\[-4ex]

{\bf ASSUMPTIONS}\\

To enable us to give moderate upper bounds on the constants $C_{XY}$,
it is necessary to make the following plausible assumption (compare
(4.3), (6.20)) throughout this section:  for all positive $\alpha$
and $\varepsilon$, and all sufficiently large $n$,

$$ t_{\alpha n}(Y) \leq (\alpha+\varepsilon)t_n(Y) \eqno{(7.3)}$$\\[-4ex]

for $Y=D, I, M, Q, R$ and $S$.  We also assume (6.6).\\

Table 7.1 gives the best known upper bounds on the constants $C_{XY}$.
Space does not permit a detailed proof of all these upper bounds,
but the main ideas of the proof are sketched below.\\

\begin{center}
TABLE 7.1 $\;\;\;\;\;\;$ Upper bounds on $C_{XY}$

\vspace*{8mm}

\begin{tabular}{|r|cccccc|}\hline
&&&&&&\\[-2ex]
        & $X=D$ & \W$I$ & \W$M$ & \W$Q$ & \W$R$ & \W$S$ \\[1ex]\hline
&&&&&&\\[-2ex]
$Y=D$   & \W1.0&\W1.0&\W2.0&\W3.0&\W2.0&\W2.0\\
$I$     & \W7.0&\W1.0&\W6.0&15.0 &14.0 &\W3.0\\
$M$     & \W4.0&\W3.0&\W1.0&\W4.5&\W5.5&\W1.0\\
$Q$     & 10.0 &\W4.0&\W6.0&\W1.0&\W5.0&\W3.0\\
$R$     & \W7.5&\W6.0&\W6.0&\W3.0&\W1.0&\W3.0\\
$S$     & \W7.5&\W5.5&\W2.0&\W7.0&\W9.0&\W1.0\\[1ex]\hline
\end{tabular}

\end{center}

\vspace*{3mm}

\underline{$C_{IM} \leq 3$}\\

Use the Newton iteration

$$x_{i+1} = x_i - x_i(a x_i - 1) \eqno{(7.4)}$$\\[-4ex]

to approximate $1/a$ using multiplications.  At the last iteration
it is necessary to compute $a x_i$ to precision $n$, but
$x_i(a x_i - 1)$ only to (relative) precision $n/2$.  Since
the order of convergence is 2, the assumptions (7.3) (with $\alpha =
\frac{1}{2}$) and (6.6) give

$$\textstyle C_{IM} \leq (1+\frac{1}{2})(1+\frac{1}{2}+\frac{1}{4}+ \ldots) 
                       = 3\;.   \eqno{(7.5)}$$\\[-4ex]

\underline{$C_{QM} \leq 4.5$}\\

Use the third-order iteration

$$\textstyle x_{i+1}= x_i - \frac{1}{2}x_i \left(\varepsilon_i - 
                \frac{3}{4}\varepsilon^2_i\right) \eqno{(7.6)}$$\\[-4ex]

where

$$ \varepsilon_i = a x^2_i -1 \eqno{(7.7)}$$\\[-4ex]

to approximate $a^{-\frac{1}{2}}$.  At the last iteration it is
necessary to compute $a x^2_i$ to precision $n$, $\varepsilon^2_i$
to precision $n/3$, and 
$x_i\left(\varepsilon_i - \frac{3}{4}\varepsilon^2_i\right)$ to precision
$2n/3$.  Thus

$$ \textstyle C_{QM} \leq (2+\frac{1}{3}+\frac{2}{3})(1+\frac{1}{3}+\frac{1}{9}
     + \ldots ) = \frac{9}{2}\;. \eqno{(7.8)}$$\\[-4ex]

Note that this bound is sharper than the bound $C_{QM}\leq 5$
which may be obtained from the second-order iteration

$$ \textstyle x_{i+1}= x_i - \frac{1}{2}x_i\varepsilon_i\;. \eqno{(7.9)}$$\\[-4ex]

\underline{$C_{RD} \leq 2$}\\

Use Newton's iteration

$$\textstyle x_{i+1}= \frac{1}{2}(x_i + a/x_i) \eqno{(7.10)} $$\\[-4ex]

to approximate $\sqrt{a}$.\\

\underline{$C_{MS} \leq 2$} \\

This follows from (3.19) and our assumptions.\\
  
\underline{$C_{IS} \leq 5.5$}\\

Use the third-order iteration

$$ x_{i+1} = x_i - x_i \left( \varepsilon_i - \varepsilon^2_i\right)
                        \eqno{(7.11)}$$\\[-4ex]

where

$$ \varepsilon_i = ax_i -1 \eqno{(7.12)}$$\\[-4ex]

to approximate $1/a$.\\

\underline{$C_{QS} \leq 7$}\\

Use the third-order iteration (7.6).\\

\underline{$C_{SI} \leq 3$}\\

{From} the proof of Lemma 3.7,

$$ t_{n/3}(S) \leq t_n (I) + O(n)\;. \eqno{(7.13)}$$\\[-4ex]

The result follows from the assumption (7.3) with $\alpha=3$.  (This
is the first time we have used (7.3) with $\alpha>1$.  The assumption is
plausible in view of the Sch\"{o}nhage-Strassen bound (6.5).)
Upper bounds on $C_{SQ}$ and $C_{SR}$ follow similarly.\\

\underline{$C_{MI} \leq 6$}\\

This follows from (7.1) and our bounds on $C_{MS}$ and $C_{SI}$.
Similarly for the bounds on $C_{MQ}$, $C_{MR}$ and $C_{RI}$.\\

\underline{$C_{QR} \leq 3$}\\

Use the identity

$$ a^{-\frac{1}{2}} = \frac{1}{\lambda}
\left( \sqrt{a+\lambda} -
\sqrt{a - \lambda} \right) +
O\left(\lambda^2/a^{5/2} \right)\;, \eqno{(7.14)}$$\\[-4ex]

where $\lambda$ is a power of 2 such that

$$ 2^{-n/3-1} \leq \lambda/a \leq 2^{1-n/3}\;. \eqno{(7.15)}$$\\[-4ex]

\pagebreak

Thus

$$ t_{2n/3}(Q) \leq 2t_n(R) +O(n)\;, \eqno{(7.16)}$$\\[-4ex]

and the result follows from (7.3).\\

\underline{$C_{DR} \leq 7.5$}\\

Use the identity

$$ b/a = \frac{1}{\lambda}
\left( \sqrt{a^2 + \lambda b} -
\sqrt{a^2 - \lambda b} \right) +
O\left( \lambda^2 b^3/a^5 \right)\;, \eqno{(7.17)}$$\\[-4ex]

where $\lambda$ is a power of 2 such that (for $b \neq 0$)

$$ 2^{-n/3-1} \leq \lambda b/a^2 \leq 2^{1-n/3}\;. \eqno{(7.18)}$$\\[-4ex]

Thus

$$ t_{2n/3}(D) \leq t_n(S) + 2t_n(R) + O(n)\;, \eqno{(7.19)}$$\\[-4ex]

and the result follows.\\

\underline{$C_{IR}\leq 6$}

$$ a^{-1} = (a^2)^{-\frac{1}{2}}\;, \eqno{(7.20)}$$\\[-4ex]

so

$$ C_{IR} \leq C_{SR} + C_{QR} \leq 6\;. \eqno{(7.21)}$$\\[-4ex]

The bound on $C_{IQ}$ also follows from (7.20), and then the bound
on $C_{RQ}$ follows from\\
$a^{\frac{1}{2}} = \left(a^{-1}\right)^{-\frac{1}{2}} $. 


\section{Comparison of some $mp$ methods for nonlinear equations}

In this section, we briefly consider methods for finding 
multiple-precision solutions of non-linear equations of the form

$$ f(x) = 0, \eqno{(8.1)}$$\\[-4ex]

where $f(x)$ can be evaluated for any $x$ in some domain.  Additional 
results are given in $[38]$.\\

There are many well-known results on the efficiency of various methods 
for solving (8.1), e.g., Hindmarsh $[10]$, Ostrowski $[20]$, Traub
$[27]$ and the references given in Section 1, but the results are only 
valid if arithmetic operations (in particular the evaluation of
$f(x), f'(x)$ etc.) require certain constant times.  The examples 
given below demonstrate that different considerations are relevant when
multiple-precision arithmetic of varying precision is used.\\

For simplicity, we restrict attention to methods for finding a simple
zero $\zeta$ of $f$ by evaluating $f$ at various points.  We assume
that $f$ has sufficiently many continuous derivatives in a 
neighbourhood of $\zeta$, but the methods considered do not require
the evaluation of these derivatives.\\

Since $f(x)$ is necessarily small near $\zeta$, it is not reasonable
to assume that $f(x)$ can be evaluated to within a small {\it relative\/}
error near $\zeta$.  In this section, an evaluation of $f$ ``with
precision $n$'' means with an {\it absolute\/} error of order
$2^{-n}$.  We suppose that such an evaluation requires time
$w(n) = t_n(E(f))$, where

$$ w(cn) \sim c^{\alpha}w(n) \eqno{(8.2)}$$\\[-4ex]

for some constant $\alpha > 1$ and all positive $c$.  Since $\alpha > 1$,
the bound (6.5) and condition (8.2) give

$$ \lim_{n \rightarrow \infty} t_n(M)/w(n) = 0\;, \eqno{(8.3)}$$\\[-4ex]

so we may ignore the time required for a fixed number of multiplications
and divisions per iteration, and merely consider the time required for
function evaluations.  Our results also apply if $\alpha = 1$, so long
as (8.3) holds.  (For example, the evaluation of $\exp(x)$ by the
method of Corollary 6.1 requires time
$w(n) \sim c_{37}n(\log(n))^3\log\log(n)$, which satisfies (8.2)
with $\alpha = 1$, and also satisfies (8.3).)\\

\begin{definition} \rm
If an $mp$ zero-finding method requires time $t(n) \sim C(\alpha)w(n)$ to
approximate $\zeta \neq 0$ with precision $n$, where $w(n)$ and
$\zeta$ are as above, then $C(\alpha)$ is the {\it asymptotic constant\/}
of the method.  (Not to be confused with the asymptotic error constant as
usually defined for fixed-precision methods $[2]$.)
\end{definition}

Given several $mp$ methods with various asymptotic constants, it is 
clear that the method with minimal asymptotic constant is the fastest 
(for sufficiently large $n$).  The method which is fastest may depend on
$\alpha$, as the following examples show.\\

{\bf DISCRETE NEWTON $mp$ METHODS}\\

Consider iterative methods of the form

$$ x_{i+1} = x_i - f(x_i)/g_i\;, \eqno{(8.4)}$$\\[-4ex]

where $g_i$ is a finite-difference approximation to $f'(x_i)$.  If
$\varepsilon_i = |x_i - \zeta|$ is sufficiently small,
$f(x_i)$ is evaluated with absolute error 
$O\left(\varepsilon^2_i\right)$, and

$$ g_i = f'(x_i) + O(\varepsilon_i)\;, \eqno{(8.5)}$$\\[-4ex]

then 

$$|x_{i+1}-\zeta| = O\left(\varepsilon^2_i\right)\;, 
                                             \eqno{(8.6)}$$\\[-4ex]

so the method has order (at least) 2.\\

The simplest method of estimating $f'(x_i)$ to sufficient accuracy is to use
the one-sided difference

$$ g_i= \frac{f(x_i+h_i)-f(x_i)}{h_i}\;, \eqno{(8.7)}$$\\[-4ex]

where $h_i$ is of order $\varepsilon_i$, and the evaluation of
$f(x_i+h_i)$ and $f(x_i)$ are performed with an absolute error
$O\left(\varepsilon^2_i\right)$.  Thus, to obtain $\zeta$ to precision
$n$ by this method $(N_1)$, we need two evaluations of $f$ to
precision $n$ (at the last iteration), preceded by two evaluations
to precision $n/2$, etc.  (The same idea is used above, in the proof
of Theorem 5.2.)  The time required is

$$ t(n) \sim 2w(n) + 2w(n/2) + 2w(n/4) + \ldots \;.\eqno{(8.8)}$$\\[-4ex]

Thus, from (8.2) and Definition 8.1, the asymptotic constant is

$$ C_{N_1}(\alpha) = 2(1+2^{-\alpha}+2^{-2\alpha} + \ldots) 
                = 2/(1-2^{-\alpha}) \;.  \eqno{(8.9)}$$\\[-4ex]

Since

$$ 2 < C_{N_1}(\alpha) \leq 4 \; , \eqno{(8.10)}$$\\[-4ex]

the time required to solve (8.1) to precision $n$ is only a small
multiple of the time required to evaluate $f$ to the same precision.
The same applies for the methods described below.\\

Using (8.7) is not necessarily the best way to estimate $f'(x_i)$.
Let $p$ be a fixed positive integer, and consider estimating $f'(x_i)$
by evaluating $f$ at the points 
\[
x_i - \lfloor p/2 \rfloor h_i, x_i -(\lfloor p/2\rfloor -1)h_i,\ldots,
x_i + \lceil p/2\rceil h_i\;.
\]  
(The points need not be equally spaced so long
as their minimum and maximum separations are of order $h_i$.)  Let
$g_i$ be the derivative (at $x_i$) of the Lagrange interpolating 
polynomial agreeing with $f$ at these points.  Since estimates $f'(x_i)$
with truncation error $O\left(h^p_i\right)$, we need $h_i$ of order
$\varepsilon^{1/p}_i$.  Then, to ensure that (8.5) holds, the function 
evaluations at the above points must be made with absolute error
$O\left(\varepsilon^{1+1/p}_i\right)$.  Thus to obtain $\zeta$ to
precision $n$ by this method $(N_p)$ we need one evaluation $f$ to
precision $n$ and $p$ evaluations to precision $n(1+1/p)/2$, preceded
by one evaluation precision $n/2$ and $p$ to precision $n(1+1/p)/4$,
etc.  The asymptotic constant is

$$C_N(p,\alpha)=\left(1+p
\left(\frac{p+1}{2p}\right)^{\alpha}\right)\Big/
(1-2^{-\alpha}) \;. \eqno{(8.11)}$$\\[-4ex]

Let

$$ C_N(\alpha) = \min_{p=1,2,\ldots}C_N(p, \alpha) \;, \eqno{(8.12)}$$\\[-4ex]

so the ``optimal $mp$ discrete Newton method'' has asymptotic constant
$C_N(\alpha)$.  From (8.11), the $p$ which minimizes $C_N(p, \alpha)$
also minimizes $p^{1/\alpha}(1+1/p)$, so the minimum for $\alpha >1$
occurs at $p=\lfloor\alpha -1\rfloor$ or $\lceil\alpha - 1\rceil$.  
In fact, $p=1$ is optimal if

$$ 1 \leq \alpha < \log(2)/\log(4/3)=2.4094 \ldots \;, \eqno{(8.13)}$$\\[-4ex]

and $p \geq 2$ is optimal if

$$ \frac{\log(1-p^{-1})}{\log(1-p^{-2})} < \alpha <
   \frac{\log(1+p^{-1})}{\log(1+1/(p(p+2)))} \;. \eqno{(8.14)}$$\\[-4ex]

The result that method $N_2$ is more efficient than method $N_1$ if
$\alpha > 2.4094 \ldots$ is interesting, for $N_2$ requires one more
function evaluation per iteration than $N_1$, and has the same order
of convergence.  The reason is that not all the function evaluations 
need to be as accurate for method $N_2$ as for method $N_1$.  
We give
below several more examples where methods with lower order and/or more 
function evaluations per iteration are more efficient than methods with
higher order and/or less function evaluations per iteration.\\

For future reference, we note that 

$$1<C_N(\alpha)\leq 4 \;, \eqno{(8.15)}$$\\[-4ex]

$$C_N(1)=4 \;, \eqno{(8.16)}$$\\[-4ex]

and

$$ C_N(\alpha) - 1 \sim e\alpha2^{-\alpha} \eqno{(8.17)}$$\\[-4ex]

as $\alpha \rightarrow \infty$.\\

{\bf A CLASS OF $mp$ SECANT METHODS}\\

It is well-known that the secant method is more efficient that the
discrete Newton method for solving nonlinear equations with 
fixed-precision arithmetic $[2,\; 20]$.  For $mp$ methods the comparison
depends on the exponent $\alpha$ in (8.2).\\
\pagebreak[4]

Let $k$ be a fixed positive integer and $p_k$ the positive real root of

$$ x^{k+1} = 1 + x^k \;. \eqno{(8.18)}$$\\[-4ex]

The iterative method $S_k$ is defined by

$$ x_{i+1} = x_i - f(x_i) \left(\frac{x_i-x_{i-k}}{f(x_i)-f(x_{i-k})}\right) \;,
                      \eqno{(8.19)}$$\\[-4ex]

where the function evaluations are performed to sufficient accuracy
to ensure that the order of convergence is at least $p_k$.  Thus,
$S_1$ is the usual secant method with order $p_1 = \frac{1+\sqrt{5}}
{2} = 1.618 \ldots; S_2, S_3$ etc.~are methods with lower orders $p_2=
1.4655 \ldots, p_3= 1.3802 \ldots$, etc.  With fixed-precision $S_1$
is always preferable to $S_2, S_3$ etc., but this is not always true 
if $mp$ arithmetic is used.\\

Suppose $i$ and $k$ fixed, $\delta > 0$ small, 
and write $\varepsilon = |x_{i-k}-\zeta|$ and $p=p_k-\delta$.  
Since the order of convergence is at least $p$, we have

$$|x_i-\zeta| = O\left(\varepsilon^{p^k}\right) \;,
                       \eqno{(8.20)}$$\\[-4ex]

$$|x_{i+1}-\zeta| = O\left(\varepsilon^{p^{k+1}}\right) \;,
                       \eqno{(8.21)}$$\\[-4ex]

$$|x_i - x_{i-k}| = O(\varepsilon) \;,
                        \eqno{(8.22)}$$\\[-4ex]

and

$$|f(x_i)| = O\left(\varepsilon^{p^k}\right) \;.
                         \eqno{(8.23)}$$\\[-4ex]

For the approximate evaluation of the right side of (8.19) to give
order $p$, the absolute error in the evaluation of $f(x_i)$ must be
$O\left(\varepsilon^{p^{k+1}}\right)$, and the relative error in the
evaluation of $(f(x_i)-f(x_{i-k}))/(x_i-x_{i-k})$ must be
$O\left(\varepsilon^{p^{k+1}-p^k}\right)$, so the absolute error
in the evaluation of $f(x_{i-k})$ must be
$O\left(\varepsilon^{p^{k+1}-p^k+1}\right)$.  From (7.18), for $\delta$
sufficiently small,

$$ p^{k+1}-p^k+1>p \;, \eqno{(8.24)}$$\\[-4ex]

so the evaluation of $\zeta$ to precision $n$ by method $S_k$ requires
evaluations of $f$ to precision $n, n/p, n/p^2,$ $ \ldots, n/p^{k-1},
2n/p^{k+1}, 2n/p^{k+2}$, etc.  Thus, the asymptotic constant is

$$ C_S(k, \alpha)= 1+p^{-\alpha}+ \ldots + p^{(1-k)\alpha} +
  (2p^{-(k+1)})^{\alpha}(1+p^{-\alpha}+ \ldots) $$

$$\hspace*{-30mm} = \frac{1-p^{-k\alpha}+(2p^{-(k+1)})^\alpha}
         {1-p^{-\alpha}} \;,
                         \eqno{(8.25)}$$\\[-4ex]

where (after letting $\delta \rightarrow 0$) $p=p_k$ satisfies (8.18).\\

We naturally choose $k$ to minimize $C_S(k, \alpha)$, giving the
``optimal $mp$ secant method'' with asymptotic constant

$$ C_S(\alpha) = \min_{k=1,2,\ldots} C_S(k,\alpha) \;.
                               \eqno{(8.26)}$$\\[-4ex]

The following lemmas show that the optimal secant method is
$S_1$ if $\alpha < 4.5243 \ldots$, and $S_2$ if 
$\alpha > 4.5243 \ldots$\\ 

\begin{lemma}
$$ C_S(k, 1) = 3 + p^k_k - p_k \;. \eqno{(8.27)}$$\\[-4ex]
\end{lemma}

{\bf Proof.}  Easy from (8.18) and (8.25).\\

\begin{lemma}

$$ C_S(k, \alpha) - 1 \sim \left\{ 
    {(3-\sqrt{5})^{\alpha}\; {\it if}\; k =1\;,}\atop
    {\hspace*{-.5mm}p^{-\alpha}_k \hspace*{10.5mm}{\it if}\; k \geq 2\;,}
                                     \right. \eqno{(8.28)}$$\\[-4ex]

as $\alpha \rightarrow \infty$.
\end{lemma}

{\bf Proof.}  From (8.25),

$$ C_S(k, \alpha) - 1 \sim p^{-\alpha}_k - p^{-k\alpha}_k +
\left(2p^{-(k+1)}_k\right)^\alpha \eqno{(8.29)}$$\\[-4ex]

as $\alpha \rightarrow \infty$.  If $k \geq 2$ then, from (8.18),

$$ p^{k}_k = p^{-1}_k + p^{k-1}_k \geq p^{-1}_k + p_k > 2 \;, 
                          \eqno{(8.30)}$$\\[-4ex]

so

$$ p^{-1}_k > 2p^{-(k+1)}_k \;. \eqno{(8.31)}$$\\[-4ex]

Thus, the result for $k \geq 2$ follows from (8.29).  The result for
$k=1$ also follows from (8.29), for $2p^{-2}_1 = 3 - \sqrt{5}$.\\

\begin{lemma}

$$ C_S(\alpha) = \left\{ 
{C_S(1, \alpha) \;\;{\it if}\;\; 1 \leq \alpha \leq \alpha_0 \;,}\atop
{\hspace*{-7mm}C_S(2, \alpha) \;\;{\it if}\;\; \alpha \geq \alpha_0 \;,}
               \right.
                     \eqno{(8.32)}$$\\[-4ex]

where $\alpha_0 = 4.5243 \ldots$ is the root of

$$ C_S(1,\alpha_0) = C_S(2,\alpha_0) \;. \eqno{(8.33)}$$\\[-4ex]
\end{lemma}

{\bf Proof.}  The details of the proof are omitted, but we note that
the result follows from Lemmas 8.1 and 8.2 for 
(respectively)	
small and large values
of $\alpha$.\\

{From} (8.25), $C_S(k, \alpha)$ is a monotonic decreasing function of
$\alpha$, so the same is true of $C_S(\alpha)$.  Thus, from Lemmas
8.1, 8.2 and 8.3,

$$ 1 < C_S(\alpha) \leq 3 \;, \eqno{(8.34)}$$\\[-4ex]

$$ \hspace*{4mm} C_S(1) = 3 \;, \eqno{(8.35)}$$\\[-4ex]

and

$$ C_S(\alpha) - 1 \sim p^{-\alpha}_2 = (0.6823 \ldots)^\alpha
                        \eqno{(8.36)}$$\\[-4ex]

as $\alpha \rightarrow \infty$.  Comparing these results with (8.15)
to (8.17), we see that the optional $mp$ secant method is more efficient
than the optimal $mp$ discrete Newton method for small $\alpha$, but
less efficient for large $\alpha$.  (The changeover occurs at $\alpha =
8.7143 \ldots)$\\

{\bf AN $mp$ METHOD USING INVERSE QUADRATIC INTERPOLATION}\\

For fixed-precision arithmetic the method of inverse quadratic interpolation
$[2]$ is slightly more efficient than the secant method, for it has
order $P_Q = 1.8392 \ldots > 1.6180 \ldots$, and requires the same 
number (one) of function evaluations per iteration.  For $mp$
arithmetic, it turns out that inverse quadratic interpolation $(Q)$
is always more efficient than the secant method $S_1$, but it is
less efficient than the secant method $S_2$ if $\alpha > 5.0571 \ldots$\\
\pagebreak

Since the analysis is similar to that for method $S_1$ above, the
details are omitted.  The order $p_Q$ is the positive real root of

$$ x^3 = 1 + x + x^2 \;. \eqno{(8.37)}$$\\[-4ex]

For brevity, we write $\sigma = 1/p_Q = 0.5436 \ldots$\\

To evaluate $\zeta$ to precision $n$ by method $Q$ requires evaluations of
$f$ to precision $n$, $(1-\sigma +\sigma^2)n$, and $\sigma^j(1-\sigma 
- \sigma^2 + 2\sigma^3)n$ for $j = 0, 1, 2, \ldots$ Hence, the
asymptotic constant is

$$    C_Q(\alpha) = 1 + (1-\sigma + \sigma^2)^\alpha 
                   + (1-\sigma - \sigma^2 
                    + 2\sigma^3)^\alpha/(1-\sigma^\alpha)$$
$$\hspace*{-11mm} = 1+ (1-\sigma + \sigma^2)^\alpha + (3\sigma^3)^\alpha/
                     (1-\sigma^\alpha) 
                    \eqno{(8.38)}$$\\[-4ex]

from (8.31).  Corresponding to the results (8.15) to (8.17) and
(8.34) to (8.36), we have that $C_Q(\alpha)$ is monotonic decreasing,

$$\textstyle 1 < C_Q(\alpha) \leq C_Q(1) = \frac{1}{2}(7-2\sigma - \sigma^2)
    = 2.8085 \ldots \;, \eqno{(8.39)}$$\\[-4ex]

and

$$ C_Q(\alpha) - 1 \sim (1-\sigma+ \sigma^2)^\alpha 
                  = (0.7519 \ldots)^\alpha
                     \eqno{(8.40)}$$\\[-4ex]

as $\alpha \rightarrow \infty$.  Method $Q$ is more efficient than the
optimal $mp$ secant method if $\alpha < 5.0571 \ldots$, and more
efficient than the optimal $mp$ discrete Newton method if $\alpha <
7.1349 \ldots$  We do not know any $mp$ method which is more efficient 
than method $Q$ for $\alpha$ close to 1.\\

{\bf OTHER $mp$ METHODS USING INVERSE INTERPOLATION}\\

Since inverse quadratic interpolation is more efficient than linear
interpolation (at least for $\alpha$ close to 1), it is natural to 
ask if inverse cubic or higher degree interpolation is even more
efficient.  Suppose $\frac{1}{2} \le \mu < 1$, and consider an inverse
interpolation method $I_\mu$ with order $1/\mu$.  In particular,
consider the method $I_\mu$ which uses inverse interpolation at
$x_i, x_{i-1}, \ldots, x_{i-k}$ to generate $x_{i+1}$, where
$k$ is sufficiently large, and the function evaluations at $x_i,
\ldots, x_{i-k}$ are sufficiently accurate to ensure that the
order is at least $1/\mu$ and, in general, no more than $1/\mu$.
(The limiting case $I_{1/2}$ is the method which uses inverse
interpolation through all previous points $x_0, x_1, \ldots, x_i$
to generate $x_{i+1}$.)\\

By an analysis similar to those above, it may be shown that the
asymptotic constant of method $I_\mu$ is

$$ C_I(\mu, \alpha) = \sum^{\infty}_{j=0} \left(s_j(\mu)\right)^\alpha \;,
                   \eqno{(8.41)}$$\\[-4ex]

where $s_0(\mu) =1$ and

$$ s_j(\mu) = \max\left[ \mu s_{j-1}(\mu), 1+j\mu^{j+1}-\mu(1-\mu^j)/
         (1-\mu)\right]  \eqno{(8.42)}$$\\[-4ex]

for $j = 1, 2, \ldots$  Space does not allow a proof of (8.41), but
related results are given in $[20,$ Appendix H$]$.  We note the 
easily verified special cases

$$ C_I\left(\frac{\sqrt{5}-1}{2},\; \alpha\right) = C_S(1, \alpha)
                        \eqno{(8.43)}$$\\[-4ex]

and

$$C_I(\sigma, \alpha) = C_Q(\alpha) \;. \eqno{(8.44)}$$\\[-4ex]
\pagebreak

The method with maximal order (see $[7]$) is $I_{1/2}$, with
asymptotic constant

$${\textstyle C_I(\frac{1}{2}, \alpha)} = \sum^{\infty}_{j=2} (j2^{1-j})^\alpha \;.
                     \eqno{(8.45)}$$\\[-4ex]

The ``optimal $mp$ inverse interpolatory method'' is the method
$I_\mu$ with $\mu(\alpha)$ chosen to minimize $C_I(\mu, \alpha)$,
so its symptotic constant is

$$ C_I(\alpha)= \min_{\frac{1}{2} \leq \mu \leq 1} C_I(\mu, \alpha) \;.
                       \eqno{(8.46)}$$\\[-2ex]

The following lemma shows that the optimal choice is $\mu = \sigma$,
corresponding to the inverse quadratic method $Q$ discussed above,
if $\alpha \leq 4.6056 \ldots$\\

\begin{lemma}
If $C_I(\alpha) = C_I(\mu(\alpha), \alpha)$ then

$$ \mu(\alpha) = \sigma = 0.5436 \ldots \;{\it if}\; 1 \leq \alpha \leq 4.6056
        \ldots \;, \eqno{(8.47)}$$\\[-4ex]

$\mu(\alpha)$ is a monotonic decreasing function of $\alpha$, and

$$\textstyle \lim_{\alpha \rightarrow \infty} \mu(\alpha) = \frac{1}{2} \;.
             \eqno{(8.48)}$$\\[-4ex]
\end{lemma}

{From} (8.39),

$$\textstyle C_I(\frac{1}{2}, \alpha) - 1 \sim \left(\frac{3}{4}\right)^\alpha
                     \eqno{(8.49)}$$\\[-4ex]

as $\alpha \rightarrow \infty$, so Lemma 8.4 shows that the optimal
inverse interpolatory is more efficient than methods $S_1$ and $Q$
(as expected), but less efficient than method $S_2$ or the optimal
discrete Newton method, for large $\alpha$.  In fact $C_I(\alpha) <
C_S(\alpha)$  for $1 \leq \alpha < 5.0608 \ldots$\\

{\bf A LOWER BOUND FOR $C(\alpha)$}\\

The following theorem shows that $C(\alpha) \geq 1$ for all useful
$mp$ methods.  The results above (e.g.~(7.17)) show that the constant
``1'' here is best possible, as methods with $C(\alpha) \rightarrow
1$ as $\alpha \rightarrow \infty$ are possible.  The minimal value
of $C(\alpha)$ for any finite $\alpha$ is an open question.\\

\begin{theorem}
If an $mp$ method is well-defined and converges to a zero of the
functions $f_1(x) = F(x) - y$ and $f_2(y) = F^{(-1)}(y) - x$, where
$x$ and $y$ are restricted to nonempty domains $D_x$ and $D_y$, and
$F$ is some invertible mapping of $D_x$ onto $D_y$ such that
$t_n(E(F))$ satisfies {\rm{(8.2)}}, then the asymptotic constant of the
method satisfies $C(\alpha) \geq 1$.
\end{theorem} \vspace*{3mm}

{\bf Proof.}  If $C(\alpha)<1$ then, by solving $f_1(x) =0$, we can
evaluate $F^{(-1)}(y)$ (for $y$ in $D_y$) in time less than
$t_n(E(F))$, for all sufficiently large $n$.  Applying the same
argument to $f_2(y)$, we can evaluate $F=(F^{(-1)})^{(-1)}$ in time
less than $t_n(E(F^{(-1)}))$.  Hence, for large $n$ we have

$$ t_n(E(F)) < t_n (E(F^{(-1)})) < t_n(E(F)) \;, \eqno{(8.50)}$$\\[-4ex]

a contradiction.  Hence, $C(\alpha) \geq 1$.\\

\begin{conjecture}
For all $mp$ methods (using only function evaluations) which are
well-defined and convergent for some reasonable class of functions
with simple zeros,

$$ C(\alpha) \geq 1/(1-2^{-\alpha}) \;. \eqno{(8.51)}$$\\[-4ex]

\end{conjecture}
\pagebreak

{\bf SUMMARY OF $mp$ ZERO-FINDING METHODS}\\

Of the methods described in this section, the most efficient are:

\begin{enumerate}
\item optimal inverse interpolation, if $1 \leq \alpha \leq 5.0608 \ldots$
      (equivalent to inverse quadratic interpolation, if $1\leq \alpha
       \leq 4.6056 \ldots)\;$;
\item optimal secant method (method $S_2$), if $5.0608 \ldots < \alpha
       \leq 8.7143 \ldots\;$;
\item optimal discrete Newton, if $8.7143 \ldots < \alpha$.
\end{enumerate}

For practical purposes, the inverse quadratic interpolation method is to 
be recommended, for it is easy to program, and its asymptotic 
$C_Q(\alpha)$ is always within 3.2\% of the least constant
for the methods above.  Numerical values of the asymptotic constants,
for various values of $\alpha$, are given to $4D$ in Table 8.1.
The smallest constant for each $\alpha$ is italicized.

\vspace*{3mm}

\begin{center}
TABLE 8.1 $\;\;\;\;\;\;$ Aysmptotic constants for various $mp$ methods

\vspace*{8mm}

\begin{tabular}{rcccccc}\hline
&&&&&&\\[-2ex]
$\alpha\W$ & $C_N(\alpha)$ & $C_S(1,\alpha)$ & $C_S(2,\alpha)$ & 
          $C_Q(\alpha)$ & $C_I(\alpha)$ & $C_I(\frac{1}{2}, \alpha)$ \\[1ex]\hline
&&&&&&\\[-2ex]

 1.0&4.0000&3.0000&3.6823&{\it 2.8085}&{\it 2.8085}&3.0000\\
 1.1&3.7489&2.8093&3.4256&{\it 2.6484}&{\it 2.6484}&2.8193\\
 1.5&3.0938&2.2987&2.7241&{\it 2.2108}&{\it 2.2108}&2.3219\\
 2.0&2.6667&1.9443&2.2209&{\it 1.8954}&{\it 1.8954}&1.9630\\
 3.0&2.1071&1.5836&1.6935&{\it 1.5586}&{\it 1.5586}&1.5856\\
 4.0&1.6988&1.3988&1.4248&{\it 1.3789}&{\it 1.3789}&1.3898\\
 5.0&1.4260&1.2860&1.2694&1.2677&{\it 1.2676}&1.2718\\
 6.0&1.2529&1.2105&{\it 1.1741}&1.1936&1.1930&1.1946\\
 7.0&1.1469&1.1573&{\it 1.1137}&1.1420&1.1410&1.1416\\
 8.0&1.0838&1.1185&{\it 1.0748}&1.1051&1.1039&1.1041\\
 9.0&{\it 1.0471}&1.0898&1.0495&1.0782&1.0770&1.0771\\
10.0&{\it 1.0262}&1.0682&1.0328&1.0584&1.0573&1.0573\\
15.0&{\it 1.0012}&1.0176&1.0043&1.0139&1.0134&1.0134\\
20.0&{\it 1.0001}&1.0046&1.0006&1.0033&1.0032&1.0032\\[1ex]\hline
\end{tabular}
\end{center}

\vspace{20mm}

{\bf NOTE ADDED IN PROOF.}  Theorem 6.2 and its corollaries may be
improved by a factor $\log(n)$, as described in $[37]$ and $[38]$.

%
%

\vspace*{\fill}
\pagebreak


\section*{References}
\smallskip

\begin{tabular}{lp{414pt}}

$[1]$& Brent, R.P. The computational complexity of iterative methods
for systems of nonlinear equations.  In {\it Complexity of Computer
Computations} (edited by R.E.~Miller and J.W.~Thatcher).  Plenum Press,
New York, 1972, 61--71.\\
$[2]$& Brent, R.P. {\it Algorithms for Minimization without Derivatives.}
Prentice-Hall, Englewood Cliffs, New Jersey, 1973.\\
$[3]$& Brent, R.P. Some efficient algorithms for solving systems of
nonlinear equations.  {\it SIAM Numer.~Anal.}~{\bf 10}, 327--344, 1973.\\
$[4]$& Brent, R.P. On the precision attainable with various 
floating-point number systems. {\it IEEE Trans.~Comp.}~{\bf C-22}, 601--607,
1973.\\
$[5]$&Brent, R.P. Error analysis of algorithms for matrix multiplication
and triangular decomposition using Winograd's identity.  
{\it Numer.~Math.}~{\bf 16}, 145--156, 1970. \\
$[6]$& Brent, R.P. 
{\it Numerical solution of nonlinear equations.} Computer Sci.\ Dept.,
Stanford University, March 1975, 189~{\it{pp}}.\\
$[7]$& Brent, R.P., Winograd, S.~and Wolfe, P. Optimal iterative
processes for rootfinding. {\it Numer.~Math.}~{\bf 20}, 327--341, 1973.\\
$[8]$& Cook, S.A.~and Aanderaa, S.O. On the minimum complexity of
functions.  {\it Trans.\ Amer.\ Math.\ Soc.}~{\bf 142}, 291--314, 1969.\\
$[9]$& Floyd, R.W. Unpublished notes.\\
$[10]$& Hindmarsh, A.C. Optimality in a class of rootfinding algorithms.
{\it SIAM J.~Numer.~Anal.} {\bf 9}, 205--214, 1972.\\
$[11]$& Hopcroft, J.E. Complexity of computer computations.  
In {\it Information Processing 74}.  North-Holland, Amsterdam, 1974,
620--626.\\
$[12]$& Karatsuba, A.~and Ofman, Y.  Multiplication of multidigit numbers
on automata (Russian). {\it Dokl.~Akad.~Nauk SSSR} {\bf 145}, 293-294, 1962.\\
$[13]$& Knuth, D.E. {\it The Art of Compter Programming\/}, Vol.~II,
{\it Seminumerical Algorithms\/}.  Addison Wesley, Reading, 
Massachusetts, 1969.  Errata and addenda:  Report CS 194, Computer 
Sci.~Department, Stanford University, 1970.\\
$[14]$& Kung, H.T. The computational complexity of algebraic numbers.
{\it SIAM J.~Numer.~Anal.\ }(to appear).\\ 
$[15]$& Kung, H.T. A bound on the multiplicative efficiency of iteration.
{\it J.~Computer \& System Sciences} {\bf 7}, 334--342, 1973.\\
$[16]$& Kung, H.T.~and Traub, J.F. Optimal order of one-point and 
multipoint iteration.  {\it J.~ACM\/} {\bf 21}, 643--651, 1974.\\
$[17]$& Kung, H.T.~and Traub, J.F. Computational complexity of one-point and 
multipoint iteration.  In {\it Complexity of Real Computations\/}
(edited by R.~Karp). Amer.~Math.~Soc., 
Providence, Rhode Island, 1974, 149--160.\\ 
$[18]$& Kung, H.T.~and Traub, J.F. Optimal order and efficiency for 
iterations with two evaluations.  Tech.~Report, Department of Computer
Science, Carnegie-Mellon University, 1973.\\
$[19]$& Morgenstern, J.  The linear complexity of computation.
{\it J.~ACM\/} {\bf 20}, 305--306 (1973).\\
$[20]$& Ostrowski, A.M. {\it Solution of Equations in Euclidean and
Banach Spaces}. Academic Press, New York, 1973.\\
$[21]$& Paterson, M.S. Efficient iterations for algebraic numbers.
In {\it Complexity of Computer Computations\/} (edited by R.E.~Miller
and J.W.~Thatcher).  Plenum Press, New York, 1972, 41--52.\\
$[22]$& Rissanen, J. On optimum root-finding algorithms.  
{\it J.~Math.~Anal.~Applics.} {\bf 36}, 220--225, 1971.\\

\end{tabular}

\begin{tabular}{lp{414pt}}
$[23]$& Sch\"{o}nhage, A.~and Strassen, V.  Schnelle Multiplikation
grosser Zahlen.  {\it Computing\/} {\bf 7}, 281--292, 1971.\\

$[24]$& Schultz, M.H.  The computational complexity of elliptic
partial differential equations.  In {\it Complexity of Computer
Computations\/} (edited by R.E.~Miller and J.W.~Thatcher).  Plenum
Press, New York, 1972, 73--83.\\
$[25]$& Shanks, D.~and Wrench, J.W.  Calculation of $\pi$ to 100,000
decimals.  {\it Math.~Comp.\ }{\bf{16}}  76--99, 1962.\\
$[26]$& Shanks, D.~and Wrench, J.W.  Calculation of $e$ to 100,000
decimals.  {\it Math.~Comp.\ }{\bf{23}}  679--680, 1969.\\
$[27]$& Traub, J.F. {\it Iterative Methods for the Solution of
Equations}.  Prentice-Hall, Englewood Cliffs, New Jersey, 1964.\\
$[28]$& Traub, J.F. Computational complexity of iterative processes.
{\it SIAM J.~Computing} {\bf 1}, 167--179, 1972.\\
$[29]$& Traub, J.F. Numerical mathematics and computer science.
{\it Comm.~ACM\/} {\bf 15}, 537--541, 1972.\\
$[30]$& Traub, J.F. Optimal iterative processes:  theorems and
conjectures.  In {\it Information Processing 71\/}. North-Holland,
Amsterdam, 1972, 1273--1277.\\
$[31]$& Traub, J.F. Theory of optimal algorithms.  In {\it Software
for Numerical Mathematics\/} (edited by D.J.~Evans).  Academic
Press, 1974.\\
$[32]$& Traub, J.F. An introduction  to some current research  in
numerical computational complexity.  Tech.~Report, Department of
Computer Science, Carnegie-Mellon University, 1973.\\
$[33]$& Winograd, S. On the time required to perform addition.
{\it J.~ACM\/} {\bf 12}, 277--285, 1965.\\
$[34]$& Winograd, S. On the time required to perform multiplication.
{\it J.~ACM\/} {\bf 14}, 793--802, 1967.\\
$[35]$& Wozniakowski, H.  Generalized information and maximal order of
iteration for operator equations.  {\it SIAM J.~Numer.~Anal\/}.
{\bf 12}, 121--135, 1975.\\
$[36]$& Wozniakowski, H.  Maximal stationary iterative methods for 
the solution of operator equations.  {\it SIAM J.~Numer.~Anal\/}.
{\bf 11}, 934--949, 1974.\\
$[37]$& Brent, R.P. Fast multiple-precision evaluation of elementary 
functions. {\it J.~ACM\/} {\bf 23}, 242--251, 1976.\\
$[38]$& Brent, R.P.  Multiple-precision zero-finding methods and
the complexity of elementary function evaluation.  In {\it Analytic
Computational Complexity\/} (edited by J.F.~Traub). Academic Press,
New York, 1975, 59--73.

\end{tabular}

\parskip 2mm
\pagebreak[4]
\section*{Postscript (September 1999)}

\subsection*{Historical Notes}

This paper was retyped (with minor corrections) in \LaTeX\
during August 1999.
It is available electronically from
\url{http://maths-people.anu.edu.au/~brent/pub/pub032.html}.
\medskip

The related paper
Brent~[38] is available electronically from
\url{http://maths-people.anu.edu.au/~brent/pub/pub028.html}
\medskip

The paper Kung~[14] appeared in {\em SIAM J.~Numer.\ Anal.\ }{\bf{12}} (1975), 89--96.
\subsection*{Sharper Results}

Some of the constants given in Table 7.1 can be improved,
e.g. $C_{DM}$, $C_{RM}$, $C_{DS}$, $C_{RS}$.
One source of improvement is given in a report by
Karp and Markstein\footnote{
Alan H.\ Karp and Peter Markstein, {\em High Precision Division and
Square Root}, HP Labs Report 93-93-42 (R.1), June 1993,
Revised October 1994. Available electronically from
{\tt http://{\lbrk}www.hpl.hp.com/{\lbrk}techreports/%
{\lbrk}93/{\lbrk}HPL-93-42.html}
}.
\medskip

For example, consider $C_{DM}$. We want to compute an $n$-bit
approximation to $b/a$. If $x_i \to 1/a$ as in (7.4) and we define
$y_i = bx_i$, then $y_i \to b/a$. Also, if $x_i$ satisfies the
recurrence~(7.4), then $y_i$ satisfies
$$	y_{i+1} = y_i - x_i(ay_i - b)\;.		\eqno(7.4') $$
Note that (7.4') is self-correcting because of the computation of
the residual $ay_i - b$. Suppose $x_i$ has
(relative) precision $n/2$. If we approximate $y_i = bx_i$ using
an $\frac{n}{2}$-bit multiplication, compute the
residual $ay_i - b$ using an $n$-bit multiplication,
then its product with $x_i$ using an $\frac{n}{2}$-bit multiplication,
we can apply~(7.4') to obtain $y_{i+1}$ with relative precision~$n$.
Assuming $x_i$ is obtained in time
$\sim 3M(n/2) \sim\frac{3}{2}M(n)$ (see (7.5)),
the time to obtain $y_{i+1}$ is $\sim\frac{7}{2}M(n)$,
i.e. $C_{DM} \le 3.5$, which is sharper than the bound
$C_{DM} \le 4.0$ given in Table~7.1.
\medskip

Similarly, we can obtain $C_{RM} \le 4.25$, which is sharper than
the bound $C_{RM} \le 5.5$ given in Table~7.1.
If $x_i \to a^{-1/2}$ and $y_i = ax_i \to \sqrt{a}$, we compute a precision
$n/2$ approximation $x_i$ in time $\sim\frac{9}{2}M(n/2)$
as in Section~7, then apply a final
second-order iteration for
$$	y_{i+1} = y_i - x_i(y_i^2 - a)/2	\eqno(7.9') $$
(derived by multiplying~(7.9) by $a$ and using~(7.7))
to obtain a precision $n$ approximation $y_{i+1}$ to $\sqrt{a}$.
\medskip

As a corollary, the time required for an arithmetic-geometric mean
iteration~[37,38] is reduced from ${\sim}6.5M(n)$ to ${\sim}5.25M(n)$.
\subsection*{The Definition of $n$-bit Multiplication}

Our $t_n(M)$ (see Sections 1--3)
is essentially the time required to compute the
most significant $n$ bits in
the product of two $n$-bit numbers. In Brent~[38], 
$t_n(M)$ is written as $M(n)$.
A related but subtly different function is $M^{*}(n)$,
defined as the time
required to compute the full $2n$-bit product of $n$-bit
numbers\footnote{In Brent~[37] 
we (confusingly) used the notation $M(n)$ for $M^{*}(n)$.}.
Paul Zimmermann\footnote{Personal communication, 1999.} observed that
smaller constants can sometimes be obtained
in row $Y = M$ of Table 7.1
if we use $M^{*}(n)$ instead of $M(n)$.
(We denote these constants by $C_{XM^{*}}$
to avoid confusion with the $C_{XM}$ of Table~7.1.)
For example, $C_{DM^{*}} < 3.5$ and $C_{RM^{*}} < 4.25$. 
\medskip

It is an open question whether
$$M(n) \sim M^{*}(n) \;\;{\rm as}\;\; n \to \infty\;;$$
with the best available multiplication algorithms (those based on the FFT) this
is true\footnote{Similar remarks apply if we consider
computing the product of two polynomials of degree $n-1$, and ask
either for the first $n$ terms in the product or the complete product.
Although the first computation is faster (by a factor of about two) if the
classical order $n^2$ algorithms are used, it is not significantly faster
if FFT-based algorithms are used.}.
\medskip
\subsection*{Final Comments}

Daniel Bernstein\footnote{Personal communication, 1999.} observed
that the time required to compute $n$-bit square roots can be
reduced further if the model of computation is relaxed so that
redundant FFTs can be eliminated. Similar remarks apply to
division, exponentiation etc (and to operations on power series).
\medskip

In conclusion, 25 years after the paper was written (in 1974),
improvements can still be found, and the last word is
yet to be written!

\end{document}